# TRACKING FINANCIAL CRIME THROUGH CODE AND LAW: A REVIEW OF REGTECH APPLICATIONS IN ANTI-MONEY LAUNDERING AND TERRORISM FINANCING

## Mariam El Harras *, My Abdelouhab Salahddine **

*\* Corresponding author,* National School of Business and Management of Tangier, Abdelmalek Essaadi University, Tangier, Morocco
Contact details: National School of Business and Management of Tangier, Abdelmalek Essaadi University, Boulevard Moulay Rchid, Airport Road, P. O. Box 1255, 90000 Tangier, Morocco
\*\* National School of Business and Management of Tangier, Abdelmalek Essaadi University, Tangier, Morocco

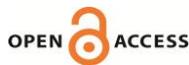

**How to cite this paper:** El Harras, M., & Salahddine, M. A. (2025). Tracking financial crime through code and law: A review of RegTech applications in anti-money laundering and terrorism financing. *Corporate Law & Governance Review, 7*(3), 73–85. https://doi.org/10.22495/clgrv7i3p7

## Abstract

Regulatory technology (RegTech) is transforming financial compliance by integrating advanced information technologies to strengthen anti-money laundering and countering the financing of terrorism (AML-CFT) frameworks. Recent literature suggests that such technologies represent more than just an efficiency tool; they mark a paradigm shift in regulation and the evolution of financial oversight (Kurum, 2023). This paper aims to provide a narrative review of recent RegTech applications in financial crime prevention, with a focus on key compliance domains. A structured literature review was conducted to examine publications between 2020 and 2024 with a thematic synthesis of findings related to customer due diligence (CDD) and know your customer (KYC), transaction monitoring, regulatory reporting and compliance automation, information sharing and cross-border cooperation, as well as cost efficiency. Findings reveal that RegTech solutions give financial institutions more responsibility for detecting and managing financial crime risks, making them more active players in compliance processes traditionally overseen by regulators. The combined use of technologies such as artificial intelligence (AI), blockchain, and big data also generates synergistic effects that improve compliance outcomes beyond what these technologies achieve individually. This demonstrates the strategic relevance of integrated RegTech approaches.

**Authors' individual contribution:** Conceptualization — M.E.H.; Methodology — M.E.H.; Resources — M.E.H.; Validation — M.E.H. and M.A.S.; Writing — Original Draft — M.E.H.; Writing — Review & Editing — M.A.S.; Supervision — M.A.S.

**Declaration of conflicting interests:** The Authors declare that there is no conflict of interest.

**Acknowledgements:** The first author sincerely acknowledges the CNRST (Centre National pour la Recherche Scientifique et Technique) in Morocco for the "PhD-Associate Scholarship — PASS" program.

## 1. INTRODUCTION

As capitalism evolved, financial systems also underwent significant transformations. The deregulation of the 1980s, particularly under the Reagan and Thatcher administrations, facilitated the internationalization of financial flows and opened the door to the expansion of financial crimes. Indeed, concealing the proceeds of criminal activities has never been easier. Criminals can now swiftly move money and goods between countries and international financial centers by using a number of procedures, which speed up the process of laundering illegal funds (Gilmour, 2020).

The concept of money laundering first emerged in the United States during the 1920s with Al Capone, who used Laundromat chains and cash-based businesses to turn funds from illegal sales into legitimate earnings (Unger, 2013). Over time, mafia producing black money through heinous crimes and extortionist activities, including the Cosa Nostra, Colombian cartels, and Japan's Yakuza, have refined their laundering methods and expanded internationally (Vernier, 2013). This global evolution has amplified the economic consequences of such illicit activities. Financial crimes severely impact both developed and developing economies (Hendriyetty & Grewal, 2017). In response, governments are putting more and more pressure on financial institutions to bolster their AML-CFT policies. This is reflected in the strict compliance obligations imposed on the sector, as illustrated by the $2 billion fine against Danske Bank for major AML failures in its Estonian branch (U.S. Securities and Exchange Commission, 2022).

As a result, technological solutions to combat financial crimes have gained prominence, particularly through RegTech, which represents the innovative use of digital tools to enhance and automate regulatory compliance, monitoring, and reporting processes in the financial sector (Arner et al., 2017). This trend aligns with findings from a Financial Action Task Force (FATF) report on digital transformation, which examines the role of technology in AML-CFT efforts and highlights that financial institutions, tech developers, and globally regulated financial technology (FinTech) companies are leading the adoption of these innovations (FATF, 2021). According to this report, these technologies will also help to reinforce efforts in AML-CFT, notably through more precise risk detection and real-time monitoring of transactions. RegTech's capabilities extend beyond automation to include enhanced risk detection, real-time transaction monitoring, as well as optimized customer vigilance, especially in scenarios involving substantial data volumes and complex regulatory frameworks. Recent studies also show that banks are relying more heavily on RegTech solutions to manage compliance risks more efficiently and reduce the operational workload associated with AML-CFT procedures (Bakhos Douaihy & Rowe, 2023).

Although financial crime and money laundering have been extensively studied, their treatment remains largely disconnected from research on RegTech, which has often been confined to its operational and technical dimensions rather than its institutional implications (Turki et al., 2020; Utami & Septivani, 2022). In order to fill this gap, this study aims to offer a structured analysis of academic and institutional literature on RegTech applications in AML-CFT, with particular attention to how integrated digital solutions contribute to risk detection and automation, as well as the redesign of compliance ecosystems. To guide this inquiry, the study addresses the following research questions:

*RQ1: How is RegTech transforming the role of financial institutions in the fight against financial crime?*

*RQ2: What are the contributions and limitations of integrated technologies in AML-CFT compliance frameworks?*

This study is grounded in the literature on regulatory technology and institutional accountability, which emphasizes the role of digital infrastructures in transforming how compliance is enacted (Becker et al., 2020; Campbell-Verduyn & Hütten, 2021; Khoury et al., 2024). RegTech is becoming a "bridge" between businesses and regulatory expectations, with a two-way flow of influence (Grassi & Lanfranchi, 2022). While regulators rely on industry-developed solutions to extend their oversight, businesses integrate regulatory logic directly into their digital systems. This mutual dependence underlines the importance of regulators not only to follow the evolution of RegTech but to actively engage with it and build internal technical expertise to respond to the sophistication of money laundering strategies (Kurum, 2023). Simultaneously, financial institutions are under growing regulatory strain in an increasingly volatile global context, which makes this study both timely and necessary.

Methodologically, we followed a structured narrative literature review, based on 33 peer-reviewed studies published between 2020 and 2024, selected through Boolean-based search strategies focused on AML-CFT technologies. This review led to the following insights. This work contributes to the growing literature on digital compliance by identifying overlooked conceptual areas and proposing a governance-focused interpretation of RegTech's role in AML-CFT frameworks. Key findings indicate that these solutions improve not only operational efficiency but also reallocate compliance responsibilities and foster technological synergies while bringing new governance challenges, particularly with regard to regulatory asymmetry and institutional capacity gaps.

The rest of this paper is structured as follows. Section 2 provides background context and reviews related academic and institutional works on RegTech adoption, especially its use in AML-CFT. Section 3 outlines the methodology of the literature review. Section 4 synthesizes key findings across five application areas: customer due diligence (CDD) and know your customer (KYC), transaction monitoring, compliance automation, information sharing, and cost efficiency. Section 5 discusses the implications and persistent challenges. Section 6 concludes by summarizing the main contributions, acknowledging limitations, and suggesting directions for future research.

## 2. LITERATURE REVIEW

### 2.1. RegTech: Revolutionizing compliance in the digital age

Technology has become a fundamental part of modern financial regulation, with the accent now being placed not on human control, but on automated systems (Zetzsche et al., 2017). This evolution was facilitated by the widespread integration of information and communication technologies (ICT) by the banking sector, which provides essential tools to meet global economic challenges (Jakšič & Marinč, 2019; Mocetti et al., 2017; Navaretti et al., 2018). This technological shift in the banking sector has catalyzed an important regulatory innovation: RegTech, a cross between "regulatory" and "technological", which the Institute of International Finance (IIF) refers to as the application of technology to improve the effectiveness and efficiency of regulatory compliance (IIF, 2016). By delivering real-time data, these technologies improve market monitoring and risk detection (Anagnostopoulos, 2018).

RegTech is no longer just about automating back-office processes. It is now seen as a strategic instrument that redistributes power and compliance responsibilities across different actors (Khoury et al., 2024). It also facilitates dynamic and data-driven oversight, allowing both regulators and institutions to transition from retrospective audits to near real-time supervision (Broby et al., 2022). As the focus is on compliance challenges and cross-border transactions, studies have revealed that the most popular and in-demand RegTech solutions are those that concentrate on AML-CFT, KYC, and anti-fraud compliance (Battanta et al., 2020; Kurum, 2023). Within these domains, challenges such as cyber identity privacy and financial crimes are expected to grow significantly in importance in the coming years (KPMG, 2022). The majority of studies in the newly emerging field of RegTech concentrate on technologies such as distributed ledgers, cloud computing, artificial intelligence (AI), big data analytics, application programming interfaces (APIs), cryptography, and biometrics, which are investigated for their potential to improve financial regulation and supervision, though their use in this area is still developing (Yang & Tsang, 2018).

Notably, both technology providers and established financial institutions have made significant strides in regulatory compliance, improving customer identity verification processes and transaction oversight. These developments allow for a more streamlined and proactive approach to meeting regulatory demands, strengthening the security and transparency of the financial sector (De Koker et al., 2019).

## 2.2. Know your customer: Beyond identity to intelligence

As indicated earlier, under financial regulation, compliance within financial institutions has become increasingly important to safeguard their reputation and the integrity of their activities. As part of this, the evolution of the banking sector has brought to the fore the urgent need for sophisticated and reliable digital identity verification systems. At the heart of this is KYC, a process that requires institutions to check the identity of their customers and collect the information needed to facilitate legitimate financial transactions (Arasa & Ottichilo, 2015).

In line with FATF Recommendation 10, financial institutions must verify the identities of their clients, identify beneficial owners, and gain a clear understanding of the purpose and nature of their business relationships (FATF, 2012/2025). As a result, different onboarding procedures, identity standards, and authentication techniques result from the various KYC needs (Arner et al., 2019).

Under AML-CFT regulations, institutions must conduct KYC screening on individuals entering into a business relationship, as well as the beneficial owner, while ensuring that they are not included on blacklists, since it is strictly forbidden to maintain business relationships with them. Research indicates that technological advancements are increasingly simplifying KYC procedures. These include the deployment of remote onboarding systems, the use of biometric technologies for secure and efficient authentication, and the integration of Blockchain solutions to accelerate and enhance the reliability of specific customer due diligence processes (Gaviyau & Sibindi, 2023; Teichmann et al., 2023). However, effective integration of these technologies, such as blockchain, depends a lot on how ready institutions are, how clear the rules are, and what digital infrastructure is already in place (Al-Smadi et al., 2023).

Building on these foundations, enhanced KYC systems are now evolving further by incorporating machine learning (ML) and big data analytics, which enable the creation of adaptive and risk-based profiles that update in real time based on client behavior (Alhajeri & Alhashem, 2023; Gandhi et al., 2024). This change means that organizations can not only verify identity, but also predict and stop any suspicious activity.

## 2.3. Real-time vigilance: The rise of automated monitoring

Transaction monitoring is a critical component of AML-CFT initiatives. It leverages information systems to detect suspicious financial activities, acting as a filter by flagging potentially fraudulent actions based on limited datasets. Nevertheless, it depends on human analysts such as compliance officers to examine the warnings and decide if the transactions that have been reported pose a danger of money laundering and terrorist financing (Chau & van Dijck Nemcsik, 2020). A local financial intelligence unit, which ensures liaison between financial institutions and regulators, must be notified when such practices are discovered and cannot be justified following a comprehensive study or inquiry.

Transaction monitoring faces a key challenge: reducing false positives. While ML is seen as a promising solution to replace traditional rule-based approaches, this technological shift is still more of an aspiration than a current reality (Oztas et al., 2024). Nevertheless, the traditional software and systems used to monitor financial transactions and to verify details of the originator and beneficiary against blacklisted entries are not sufficiently effective to either unblock or block transactions from the historical database (Alkhalili et al., 2021).

In response to these limitations, the recent literature emphasizes that real-time and automated transaction monitoring is now a foundation of modern AML-CFT compliance (Gupta et al., 2023; Garcia-Segura, 2024). Rather than conducting audits after the fact, banks are recently deploying AI tools that continuously monitor transactions and flag suspicious activities as they occur.

## 3. RESEARCH METHODOLOGY

### 3.1. Search strategy and data sources

In addition to providing empirical evidence of specific effects, a literature review can facilitate the establishment of guidelines for policy and practice, contribute to the advancement of knowledge, and, when rigorously conducted, stimulate novel ideas and potential research trajectories within a particular domain (Snyder, 2019). In the context of AML-CFT technologies, it can be particularly instrumental by offering insights that may catalyze innovation, inform policymaking, and shape future scholarly inquiries in this pivotal area. The purpose is to provide a summary of the main findings, to evaluate current knowledge, to identify gaps, and to address any ambiguities in existing research (Knopf, 2006).

A structured methodology was adopted for this literature review, with data gathered from the Scopus database. Selection was based on title, abstract, and keywords to ensure reliability and accuracy in our findings. A Boolean search was conducted using AND and OR operators to explore the literature on emerging technologies in AML-CFT. Given the high number of results, the search was limited to the following keywords (Table 1).

**Table 1.** Themes, keyword pool, and final list of selected keywords used in the systematic literature review

| Category | Keyword pool | Selected keywords |
|---|---|---|
| Financial crime and identity threats | Money laundering, financial crime, terrorism financing, KYC, digital identity | "Money laundering" OR "Financial crime*" OR "Terrorism financing" |
| Technological solutions for compliance | RegTech, FinTech, artificial intelligence, machine learning, blockchain, data mining | "RegTech" OR "FinTech" OR "Artificial Intelligence" OR "Machine Learning" OR "Blockchain" OR "Data mining" |

In order to maintain relevance, our review focuses on articles and conference papers published between 2020 and 2024. Also, only publications written in English were considered for inclusion. To uphold alignment with our review's objectives, we excluded publications that did not meet these criteria to ensure the final selection was relevant and aligned with the review's objectives. Book chapters, editorials, notes, and quick polls were all excluded since they were not pertinent to the review's focus. Additionally, studies that were off-topic or that concentrated on criminal technology rather than AML-CFT were also not included.

Other methods, such as bibliometric analysis, case studies, or field research, would have provided valuable insights by enabling trends to be mapped or the practical implementation of RegTech to be examined. These approaches, however, were less suitable for capturing the broader conceptual and governance issues addressed in this study. Therefore, a structured narrative review was considered the most appropriate.

## 3.2. Screening and selection process

After eliminating duplicates to ensure the chosen studies were unique, a preliminary filtering was performed by screening the article abstracts, which allowed the elimination of studies that were not related to the main scope of the review, in particular studies that focused principally on AML-CFT for virtual currencies using blockchain technology, as they did not correspond to our main topic. Only those papers that provided significant insight into the technologies used to combat financial crime were to be considered.

The second phase involved a detailed assessment, in full text, of the selected publications, to determine their relevance and contribution to the field. This allowed us to determine the most significant and relevant research results. After careful review of the content of each publication, 33 studies were selected for inclusion in the final dataset based on their compliance with the selection criteria (Figure 1).

**Figure 1.** PRISMA Diagram illustrating the screening and inclusion process

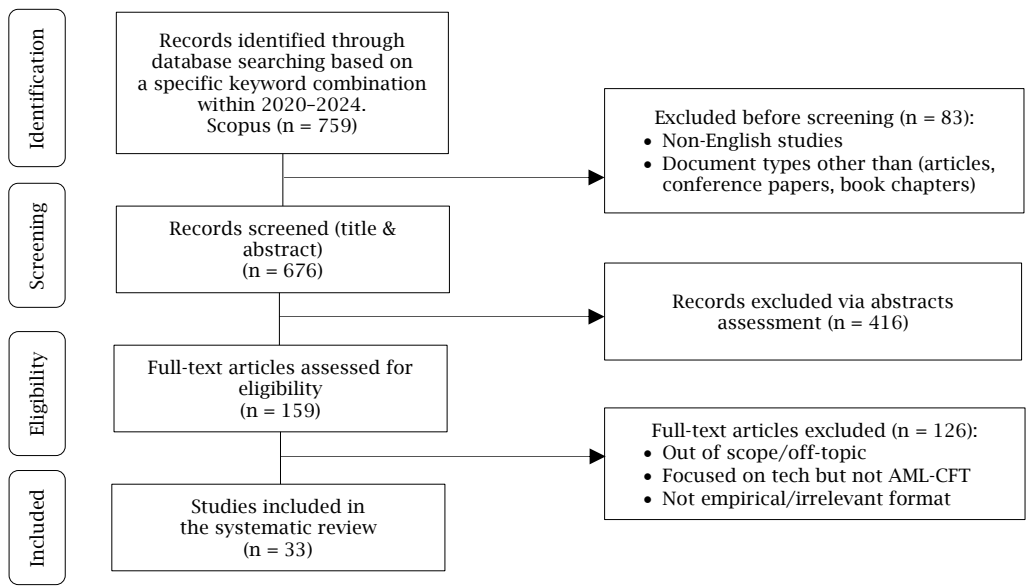

## 3.3. Epistemological and methodological approaches

Knowledge management paradigms provide structured approaches to analyzing organizational knowledge, offering both practical tools and theoretical foundations for research (Turyahikayo, 2021). Understanding these paradigms clarifies the epistemological stances and methodological choices that are likely to transform scholarly work in this field.

The positivist paradigm, which assumes that reality can be objectively measured through scientific methods and empirical evidence (Park et al., 2020), underpins most of the studies reviewed. Among the 33 articles, 64% adopt a positivist stance, employing quantitative methods such as statistical modeling, algorithmic evaluation, and performance metrics (Antwi & Hamza, 2015). For example, Al-Ababneh et al. (2024) and Sharma et al. (2024) assess the accuracy and efficiency of AI and ML tools in AML-CFT contexts.

In parallel, the interpretivist paradigm focuses on understanding meaning and experience within specific social or organizational contexts (Alharahsheh & Pius, 2020). Approximately 27% of the studies follow this approach, using qualitative methods to explore ethical concerns, implementation challenges, and user perceptions of RegTech systems. Studies by Daugaard et al. (2024) and Pavlidis (2023) show how technologies like blockchain and AI are integrated into real-world institutions. In addition, qualitative techniques such as the Delphi method, which involves iterative rounds of expert feedback to reach consensus on complex issues, are used in some studies, such as that of Kurum (2023). Others use literature review methods to synthesize existing knowledge on technologies such as Blockchain and AI, as well as to identify research gaps and suggest future research directions (Bozorgi, 2024; Smith & Tiwari, 2024;

Swain & Gochhait, 2022). This distribution (see Appendix, Table A.1) reflects a dominant orientation toward measurable outcomes, while still incorporating contextual and interpretive insights.

## 4. RESULTS

### 4.1. Publication trends

The analysis of selected publications from 2020 to 2024 reveals a clear evolution in both the volume and focus of research on RegTech technologies (see Appendix, Table A.1). This growth reflects the increasing role of these technologies in combating financial crime and ensuring regulatory compliance (see Figure 2). Early publications (2020–2021) primarily focused on establishing foundational frameworks and exploring initial applications of AI and ML in AML-CFT (Chitimira & Ncube, 2021; Couchoro et al., 2021), while more recent research (2023–2024) emphasizes practical implementations and integrated solutions, reflecting a maturing field (Lokanan, 2024a; Meiryani et al., 2023; Sharma et al., 2024; Usman et al., 2023). This progression exemplifies how academic attention has moved from theoretical potential to operational integration, and underlines the strategic relevance of RegTech in the field of financial compliance.

**Figure 2.** Evolution of the number of documents analyzed per year (2020–2024)

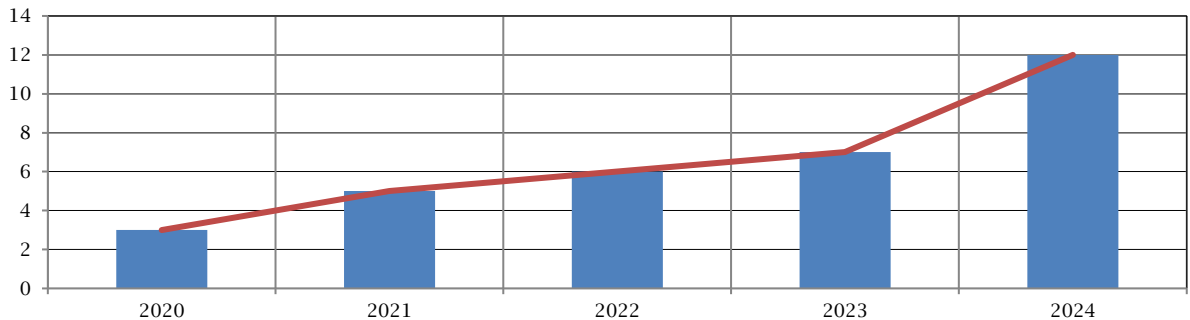

### 4.2. Technologies identified in the literature

A thorough analysis of the technologies presented in the literature uncovers the emergence of several trends that are transforming the financial compliance landscape (see Figure 3 and Appendix, Table A.1):

• AI and ML have emerged as the predominant technological solutions, with their presence in publications jumping from 70% in 2020 to 91.7% in 2024, a remarkable 21.7 percentage point increase. Applications range from neural networks (Lokanan, 2024a, 2024b; M. Raj et al., 2024) to various learning algorithms (Beketnova, 2021; Kumar et al., 2022; Mohammed et al., 2022).

• Advanced analytics also grew in significance over the period, from 22% to 33.3%. This reflects the adoption of complex data analysis methods such as pattern recognition (Canhoto, 2021;

Prisznyák, 2022) as well as graphical analysis (Garcia-Bedoya et al., 2021).

• Blockchain and distributed technologies rose from 15% to 27.8%, illustrating a move toward decentralized systems (Campbell-Verduyn & Hütten, 2021; A. Raj et al., 2023).

• In parallel, specialized technical solutions, including natural language processing (Pavlidis, 2023) and biometrics (Kurum, 2023), increased from 12% to 19.4%, addressing targeted operational needs.

The above analysis highlights a clear trend toward a multi-technology approach rather than isolated, single-technology tools. In particular, researchers are interested in technological synergetic effects that will create more comprehensive regulatory solutions. The powerful combination of AI, blockchain, and advanced analytics has considerably improved the detection of suspicious activity and the effectiveness of regulatory compliance (Daugaard et al., 2024).

**Figure 3.** Evolution of technology coverage in RegTech literature (2020–2024) in percentage

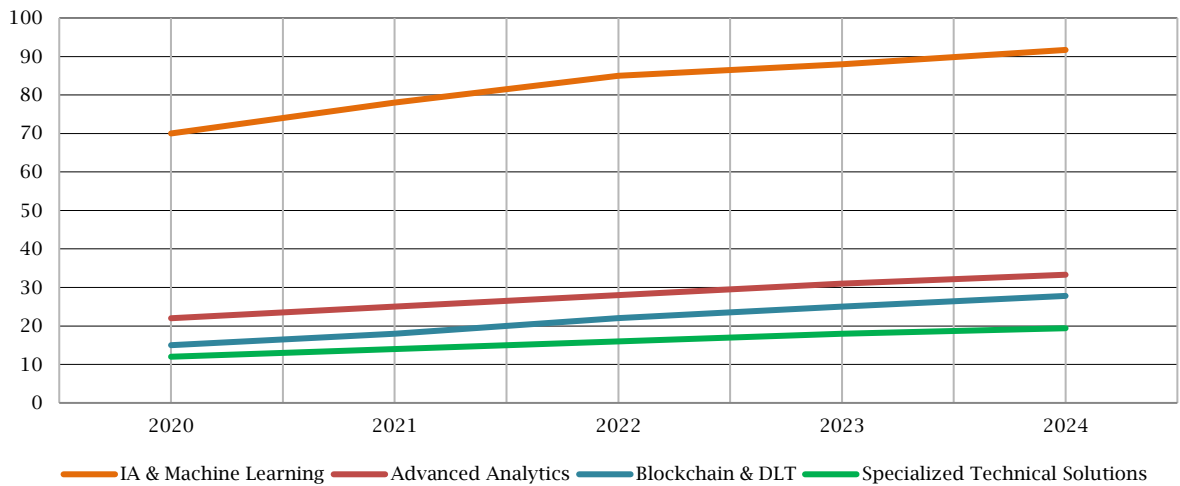

## 4.3. Main application areas of RegTech

The reviewed literature identifies five main areas where RegTech is most applied in the AML-CFT framework: CDD and KYC, transaction monitoring, regulatory reporting and compliance automation, information sharing and cross-border cooperation, and cost and time efficiency (see Appendix, Table A.2 and Figure 4). These categories reflect how RegTech integrates advanced technologies to improve compliance processes and enhance the detection of financial crime.

### 4.3.1. Customer due diligence and know your customer

Our analysis reveals that RegTech is transforming the way financial institutions approach CDD and KYC processes by leveraging advanced technologies such as Blockchain, ML, and biometric verification (Lokanan, 2019; Canhoto, 2021; Thommandru & Chakka, 2023; Bhumikapala et al., 2024). A qualitative study has demonstrated that Blockchain enables secure and reusable digital identities, which reduce redundant verifications and onboarding time (Daugaard et al., 2024). Another qualitative study based on data derived from simulated real-life transactions flagged as suspicious for money laundering in Middle Eastern banks showed that ML and artificial neural networks (ANN) algorithms assess customer risk through behavioral data and detect anomalies such as sudden spikes or irregular transactions (Kumar et al., 2022; Lokanan, 2024b). It was also demonstrated that the use of big data analytics and ML while incorporating KYC data and contextual information proves to be a highly effective strategy for identifying money laundering activities, especially with support vector machines (SVMs) (Usman et al., 2023).

**Figure 4.** Main application areas of RegTech in the AML-CFT framework

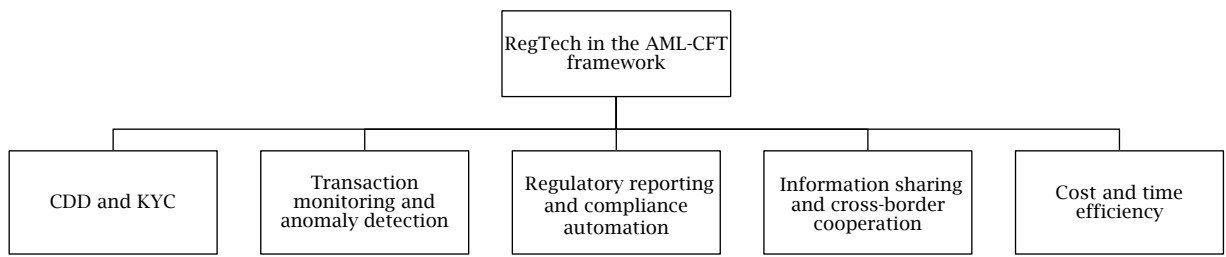

### 4.3.2. Detection of suspicious transactions and anomaly detection

Many of the articles we reviewed highlighted the significant role of RegTech in enhancing transaction monitoring. Advanced technologies capable of processing massive amounts of data are essential for the effectiveness of these systems, such as ML (Lokanan, 2024b; Mbiva & Correa, 2024; Sharma et al., 2024), AI (Garcia-Bedoya et al., 2021; Chitimira et al., 2024), and graph-based technologies (Kurshan & Shen, 2020; Huong et al., 2024). These technologies enable real-time analysis of large datasets, allowing for the detection of suspicious activities and fraud patterns.

It has been shown that reducing false alerts was one of the main contributions of emergent technologies. While traditional systems rely on rule-based scenarios that generate a large percentage of false alerts, creating a huge workload for compliance officers, those traditional rule-based manual methods have become inefficient and subject to false alarms. A study by Al-Ababneh et al. (2024) reports a 30% reduction in false positives and 25% improvement in high-risk detection using AI models over traditional rules. A ML model tested at DNB Bank in Norway has also contributed to the reduction of manual reviews by 51% and detected 80% of suspicious cases (Jullum et al., 2020), while similar research highlighted isolation forest for its real efficiency in identifying global transaction anomalies and rapidly processing suspicious activities which makes it particularly valuable for large scale financial monitoring applications (Oztas et al., 2022). Hybrid locality outlier factor-isolation forest (LOF-IF) algorithm models identify both global and local anomalies (Mbiva & Correa, 2024). In addition to cognitive and analytical technologies, blockchain provides traceable, immutable records, reducing review time (Daugaard et al., 2024).

### 4.3.3. Regulatory reporting and compliance automation

Regulatory reporting, as a fundamental part of financial institutions' obligations in AML-CFT, involves submitting financial data to regulatory authorities in order to ensure compliance with laws and regulations. In the Indonesian banking sector, for example, a survey of 160 professionals across eight banks measured three key dimensions: e-KYC systems, transaction monitoring capabilities, and operational efficiency. Results show banking professionals strongly endorse RegTech's effectiveness in enhancing risk identification and streamlining regulatory processes (Meiryani et al., 2023).

ML accelerates the process of uncovering illegal activities and non-compliant practices, and enables regulators to catch financial institutions that are under suspicion at an early stage. As a result, reactivity is improved, surveillance is strengthened, and more effective application of financial regulations is achieved (Beketnova, 2021). This shift supports a broader trend; Institutions increasingly rely on RegTech for interpreting AML-CFT policies (Kurum, 2023), with AI models being regularly retrained in order to adapt to new and evolving financial crime typologies (Al-Ababneh et al., 2024).

### 4.3.4. Information sharing and cross-border cooperation

With the complexity of international financial movements and the growing level of sophistication of criminal activity continuing to increase, a unilateral approach on the part of individual states is no longer adequate. In this context, as highlighted by Pavlidis (2023), there is a need for national and international cooperation to ensure the effective use of AI in AML-CFT, with collaboration between local authorities and global alignment of regulations and standards across sectors. Building on this, a qualitative study demonstrated that RegTech solutions enhance transparency and efficiency in cross-border operations. These tools, institutionalized under international regulatory pressure, improve AML-CFT efforts while facilitating global transactions (Bakhos Douaihy & Rowe, 2023).

In parallel, Blockchain technology is highly valued for its role in compliance, as blockchain-based networks simplify international flows, enhance traceability, and reduce fraud risks (A. Raj et al., 2023; Thommandru & Chakka, 2023). It was empirically demonstrated that a multichain distributed P2P Network improves transparency and reduces delays in cross-border transactions (A. Raj et al., 2023).

### 4.3.5. Cost and time efficiency

The literature we reviewed illustrated that the financial sector has witnessed a transformative shift in compliance operations through RegTech implementation. Institutions report major savings through automation of onboarding and transaction monitoring (Pavlidis, 2023; Prisznyák, 2022). A study conducted by Gupta et al. (2022) showed how mathematical optimization in threshold fine-tuning can generate millions in savings for financial institutions while maintaining robust monitoring capabilities. This cost-effectiveness is further reinforced by a study of Daugaard et al. (2024) showing that shared blockchain-based networks lower costs for banks working within the same compliance system. RegTech is seen as a cost-effective strategy that combines accuracy with speed (Al-Ababneh et al., 2024; Kurum, 2023).

## 5. DISCUSSION

The findings of this review reveal a profound transformation in how compliance is conceived and operationalized in the financial sector. The adoption of RegTech in AML-CFT is a reflection of more than a technological evolution; it points to a change in the institutional fabric of financial compliance. As Anagnostopoulos (2018) suggests, regulatory innovation today is no longer a matter of competing interests but of co-constructed ecosystems, where banks, FinTechs, and regulators share overlapping responsibilities. RegTech operates at the center of this convergence, not as a technical layer, but as a force that changes the balance between risk management, control, and adaptability.

One of the most notable evolutions lies in the individualization of risk detection. Thanks to AI and ML, more nuanced profiling is now possible. This enables institutions to move beyond static rule-based systems. Unlike earlier frameworks, where the detection logic was imposed from outside by regulators, current RegTech tools facilitate adaptive compliance systems that learn from behavior and adjust thresholds accordingly. This shift aligns with the conceptual reframing noted by (Khoury et al., 2024), but our synthesis goes further by anchoring it in operational realities and documenting reductions of 30% in false positives (Al-Ababneh et al., 2024) and over 50% in manual verification workloads through specific algorithmic architectures.

Another key insight is the functional convergence of technologies across AML-CFT stages. While previous literature often discusses ML, blockchain, or big data in isolation, our findings show that their true potential emerges through complementary use. For instance, blockchain addresses trust and data immutability in KYC and cross-border flows, while AI ensures dynamic transaction surveillance. Together, these tools form multi-layered compliance architectures, where identity verification, anomaly detection, and regulatory reporting are interconnected as described by Pavlidis (2023). This perspective also complements the work of Firmansyah and Arman (2022), whose architectural mapping highlights how technologies like AI and blockchain can support distinct compliance functions. From these observations, our findings show that these technologies do not operate in isolation, but rather as interdependent layers within AML-CFT workflows.

A third transformation relates to institutional responsibility. As technologies evolve, so does the distribution of compliance duties. Whereas the traditional model placed the regulator at the center, RegTech tools empower institutions to self-regulate in real time and faster than supervisory bodies can react. This evolution raises concerns about regulatory asymmetry, where the tools and expertise lie more with the private sector than with public oversight bodies. While Rafiq and Sohail (2023) touch only by pointing to the challenges regulators face in keeping pace with innovation, our review illustrates how institutions not only adopt technology but also dynamically transform the logic of compliance through tool design, data control, and internal rule calibration.

The growing autonomy of RegTech systems raises governance and oversight challenges, especially in contexts where regulatory frameworks lag behind technological capabilities (Kurum, 2023) warns of such gaps, and our review reinforces this concern: the speed at which AI models evolve often exceeds that of policy cycles, which creates a risk of regulatory mismatch. This is particularly problematic in smaller institutions or jurisdictions where the resources to implement, audit, and adapt these systems remain limited.

Finally, the promise of cost and time efficiency must be approached critically. While the literature reports substantial savings (Daugaard et al., 2024; Gupta et al., 2022), these benefits are not uniformly distributed. Smaller institutions may face barriers to adoption due to resource constraints, while the integration of RegTech requires not just financial investment but also factors often underexamined in current research, namely, genuine organizational change, talent development, as well as cultural adaptation.

# 6. CONCLUSION

While criminals consistently exploit weaknesses in traditional financial systems and stay one step ahead of conventional compliance measures (Kurum, 2023), our findings lead to three central conclusions regarding the evolving role of RegTech in financial crime prevention.

First, the progressive incorporation of advanced technologies, including AI, ML, and big data, into core AML-CFT compliance functions is changing the status of financial institutions from passive rule-followers into active risk managers. This transformation is particularly evident in areas like KYC and transaction monitoring where dynamic systems are replacing static rule-based models. In fact, these technologies do not merely support compliance processes but they are redefining which transforms financial institutions into major participants in the detection and management of financial crime risks, traditionally the domain of regulatory authorities.

Second, the integration of multiple technologies produces stronger results than isolated implementations. AI improves detection accuracy and reduces false alerts, while blockchain enhances secure identity verification and enables cross-border cooperation. However, realizing these synergies requires institutional capacity and strategic coordination, which suggests a future model based on ecosystem thinking rather than fragmented tools. This transformation raises important ethical and governance challenges. While RegTech enhances monitoring capabilities, it also redistributes responsibility across actors and systems which creates ambiguity around accountability. Many institutions lack adequate regulatory oversight mechanisms for algorithmic decisions. This becomes even more critical when technologies are deployed across borders which is challenging the coherence of international regulatory frameworks and data governance standards. In addition, smaller establishments may find it difficult to access and implement these advanced systems, which in turn exacerbates asymmetries between jurisdictions in terms of regulatory capacity and technological readiness.

The study also recognizes an important limitation, namely that while it draws on 33 recent studies to map functional impacts and trends, it is limited by the lack of direct empirical evaluation of real-world implementations. Much of the available data remains theoretical or simulation-based, especially concerning algorithmic performance and interoperability across systems. Additional investigation is needed to assess the effectiveness of RegTech adoption in practice. Studies focusing on institutional adaptation, regulatory capacity development, and the embedding of ethical considerations into system design will be of paramount importance. In parallel, the evolving nature of financial crime threats must also be considered. Cryptocurrencies, in particular, have become a growing concern. The literature underscores the central role that crypto-assets play in facilitating money laundering activities and highlights how their decentralized and often unregulated nature undermines the effectiveness of current compliance systems (Guidara, 2022; Leuprecht et al., 2023). As noted by Prendi et al. (2023, p. 90), "electronic money is the future of currencies", but its growth must be accompanied by proper infrastructure and oversight.

These findings carry both theoretical and practical implications. The paper's theoretical perspective calls for a rethinking of traditional compliance and regulatory frameworks. RegTech is more than just a toolkit; it represents a new ecosystem that merges technology, regulation, and governance. For example, blockchain solutions question the central role of regulators by shifting the compliance paradigm toward distributed decision-making. This evolution calls for updated frameworks that account for the convergence of regulation and technological innovation. From a practical standpoint, RegTech has become indispensable for institutions operating in increasingly complex and rapidly changing regulatory environments. Tailored solutions are needed for different compliance functions, alongside safeguards to ensure data security, avoid algorithmic bias, and manage integration with legacy systems (M. Raj et al., 2024). For regulators, RegTech enhances control through real-time monitoring and information sharing, improving the detection of suspicious activities and international cooperation. As adoption grows, recruiting technical specialists becomes as critical as the technology itself, enabling regulators to effectively manage the transformation revolutionizing AML-CFT compliance processes (Kurum, 2023). Establishing harmonized standards and universal data formats would support interoperability, while regulatory sandboxes could foster innovation (Grassi & Lanfranchi, 2022).

To support future research directions, this study suggests several priority areas. To expand current knowledge, deepen understanding of financial crime prevention and compliance, respond effectively to the fast-evolving challenges in these areas, and harness the full transformative potential of RegTech to foster more secure, transparent, and effective financial systems, future research should address the following critical questions:

• How can collaboration between regulators, financial institutions, and technology providers be strengthened? What models of partnership and incentive structures are most effective?

• What ethical considerations should guide the use of decentralized identity systems? How can blockchain innovations be aligned with privacy and data protection principles?

• How should RegTech evolve to address emerging financial crime threats such as those related to cryptocurrencies? Which technological adaptations are most promising?

• What role can RegTech play in supporting SDG 16? How can it help combat corruption and reduce illicit financial flows at scale?

These questions highlight persistent gaps in the literature and offer a concrete agenda for advancing research on RegTech's evolving role in AML-CFT.

# APPENDIX

**Table A.1.** Overview of the literature (Part 1)

| No. | Author | Year | Research paradigm | Method | Technologies mentioned |
|---|---|---|---|---|---|
| 1 | Al-Ababneh et al. | 2024 | Positivism | Quantitative | Artificial intelligence<br>Machine learning<br>Supervised and unsupervised learning<br>Neural networks |
| 2 | Mbiva and Correa | 2024 | Positivism | Quantitative | Machine learning<br>Unsupervised machine learning<br>Isolation forest<br>Local outlier factor<br>Ensemble outlier detection algorithm |
| 3 | Daugaard et al. | 2024 | Interpretivism | Qualitative | Blockchain technology<br>Distributed ledger technology<br>Artificial intelligence<br>Quantum computing<br>Quantum machine learning |
| 4 | Smith and Tiwari | 2024 | Positivism | Conceptual/Literature review | Blockchain<br>Digital currencies<br>Smart contracts |
| 5 | Chitimira et al. | 2024 | Interpretivism | Qualitative | Artificial intelligence<br>Machine learning<br>Big data |
| 6 | Sharma et al. | 2024 | Positivism | Quantitative | Artificial intelligence<br>Machine learning<br>Supervised machine learning<br>Unsupervised machine learning |
| 7 | Bhumikapala et al. | 2024 | Positivism | Quantitative | Machine learning<br>Support vector machine (SVM)<br>Exact string matching (ESM) |
| 8 | Bozorgi | 2024 | Positivism | Conceptual/Literature review | Blockchain<br>Artificial intelligence |
| 9 | M. Raj et al. | 2024 | Positivism | Conceptual/Literature review | Artificial Intelligence<br>Machine Learning<br>Artificial neural networks (ANN)<br>Convolutional neural networks (CNN) |
| 10 | Huong et al. | 2024 | Positivism | Quantitative | Network graphs<br>Language modeling<br>Supervised learning<br>Random forest (RF) |
| 11 | Lokanan | 2024a | Positivism | Quantitative | Machine learning<br>Artificial neural network |
| 12 | Lokanan | 2024b | Positivism | Quantitative | Machine learning models<br>Neural network models |
| 13 | Thommandru and Chakka | 2023 | Interpretivism | Qualitative | Blockchain<br>Distributed ledger technology (DLT) |
| 14 | Bakhos Douaihy and Rowe | 2023 | Interpretivism | Qualitative | Artificial intelligence<br>Machine learning |
| 15 | Meiryani et al. | 2023 | Positivism | Quantitative | Electronic know your customer (e-KYC) |
| 16 | Kurum | 2023 | Positivism | Mixed | Artificial intelligence<br>Machine learning<br>Blockchain<br>Big data analytics<br>Biometrics<br>Cloud computing |
| 17 | A. Raj et al. | 2023 | Positivism | Mixed | Blockchain<br>Proof of work (PoW)<br>Proof of stake (PoS)<br>Decentralized finance (DeFi) |
| 18 | Pavlidis | 2023 | Interpretivism | Qualitative | Artificial intelligence<br>Machine learning<br>Natural language processing |
| 19 | Naveed et al. | 2023 | Positivism | Quantitative | Graph-based machine learning |
| 20 | Kumar et al. | 2022 | Positivism | Quantitative | Supervised machine learning algorithms |
| 21 | Gupta et al. | 2022 | Positivism | Quantitative | Linear/non-linear optimization techniques<br>Artificial intelligence |
| 22 | Oztas et al. | 2022 | Interpretivism | Qualitative | Artificial Intelligence<br>Machine Learning<br>Natural Language Processing |
| 23 | Swain and Gochhait | 2022 | Interpretivism | Qualitative | Artificial intelligence<br>Blockchain<br>Cloud computing |
| 24 | Prisznyák | 2022 | Positivism | Conceptual/Literature review | Artificial Intelligence<br>Supervised learning methods (classification, regression)<br>Unsupervised learning methods (clustering, anomaly detection)<br>Big data analytics<br>Neural networks |
| 25 | Mohammed et al. | 2022 | Positivism | Conceptual/Literature review | Machine learning<br>Deep learning |

**Table A.2.** Overview of the literature (Part 2)

| No. | Author | Year | Research paradigm | Method | Technologies mentioned |
|---|---|---|---|---|---|
| 26 | Canhoto | 2021 | Interpretivism | Qualitative | Artificial intelligence<br>Machine learning<br>Big data analytics |
| 27 | Couchoro et al. | 2021 | Positivism | Quantitative | Artificial intelligence<br>Machine learning |
| 28 | Campbell-Verduyn and Hütten | 2021 | Interpretivist | Qualitative | Blockchain |
| 29 | Beketnova | 2021 | Positivist | Quantitative | Machine learning |
| 30 | Chitimira and Ncube | 2021 | Interpretivist | Qualitative | Artificial intelligence<br>Machine learning |
| 31 | Kurshan and Shen | 2020 | Positivist | Conceptual/Literature review | Artificial intelligence,<br>Machine learning<br>Graph computing,<br>Graph neural networks |
| 32 | Jullum et al. | 2020 | Positivist | Quantitative | Machine learning,<br>XGBoost,<br>Supervised learning,<br>Statistical modeling |
| 33 | Garcia-Bedoya et al. | 2021 | Positivist | Mixed | Artificial intelligence<br>Machine learning,<br>Graph analysis |

**Table A.2.** Overview of key technologies in AML-CFT: Insights from the literature

| Main theme | Key technologies | Outcomes | Studies/Examples |
|---|---|---|---|
| CDD/KYC | Blockchain, machine learning, biometrics, big data analytics | Simplification of onboarding, reduction of redundant checks, real-time identification of money laundering risks. | • Blockchain: Secure digital identity storage (Thommandru & Chakka, 2023);<br>• Machine learning: Detection of behavioral changes (Lokanan, 2024);<br>• SVM for fraud detection (Kumar et al., 2022) |
| Transaction monitoring | Machine learning, artificial intelligence, graph-based technologies, isolation forest, LOF-IF | Reduction of false positives, real-time anomaly detection, improved accuracy of alerts. | • 30% reduction in false positives with AI (Al-Ababneh et al., 2024);<br>• LOF-IF for terrorist financing detection (Mbiva & Correa, 2024);<br>• Blockchain for traceability (Daugaard et al., 2024) |
| Regulatory reporting and compliance automation | Artificial intelligence, machine learning, e-KYC | Optimization of compliance processes, continuous system updates, rapid detection of illegal activities. | • RegTech adoption in Indonesian banking (Meiryani et al., 2023);<br>• Continuous algorithm updates (Al-Ababneh et al., 2024) |
| Information sharing and cross-border cooperation | Blockchain, multichain P2P networks, International Collaboration | Increased transparency, security of cross-border transactions, alignment of international regulations | • Blockchain for international transactions (Thommandru & Chakka, 2023);<br>• Multichain P2P networks (Raj et al., 2023);<br>• International collaboration (Pavlidis, 2023) |
| Cost and time savings | Mathematical optimization, blockchain, artificial intelligence | Reduction of compliance costs, freeing up employee time for strategic tasks, operational efficiency. | • Threshold optimization for significant savings (Gupta et al., 2022);<br>• Blockchain for cost-sharing (Daugaard et al., 2024);<br>• Artificial intelligence for reducing compliance costs (Al-Ababneh et al., 2024) |